\documentstyle[12pt,epsf]{article}
\title{Thermodynamic properties
       \protect\\
       of spin-$\frac{1}{2}$ transverse $XY$ chain
       \protect\\
       with Dzyaloshinskii-Moriya interaction:
       \protect\\
       Exact solution for correlated Lorentzian disorder}
\author{Oleg Derzhko$^\dagger$ and Johannes Richter$^\ddagger$\\
\small   {\em {$^\dagger$Institute for Condensed Matter Physics}}\\
\small   {\em {1 Svientsitskii St., L'viv-11, 290011, Ukraine}}\\
\small   {\em {$^\ddagger$Institut f\"{u}r Theoretische Physik,
               Universit\"{a}t Magdeburg}}\\
\small   {\em {P.O. Box 4120, D-39016 Magdeburg, Germany}}}

\date{\today}

\begin{document}

\maketitle

\begin{abstract}
We extend the consideration of
the spin-$\frac{1}{2}$ transverse $XY$ chain
with correlated Lorentzian disorder
(Phys. Rev. B {\bf 55,} 14298 (1997))
for the case of additional 
Dzya\-lo\-shin\-skii-Moriya interspin interaction.
It is shown how the averaged density of states can be calculated exactly.
Results are presented for the density of states and the transverse
magnetization.

\end{abstract}

\vspace{2mm}


\vspace{1mm}


\vspace{0mm}

\noindent
{\bf Postal addresses:}\\
[5pt]
{\em
Dr. Oleg Derzhko (corresponding author)\\
Institute for Condensed Matter Physics\\
1 Svientsitskii St., L'viv-11, 290011, Ukraine\\
Tel: (0322) 42 74 39\\
Fax: (0322) 76 19 78\\
E-mail: derzhko@icmp.lviv.ua\\
[3pt]
Prof. Johannes Richter\\
Institut f\"{u}r Theoretische Physik, Universit\"{a}t Magdeburg\\
P.O. Box 4120, D-39016 Magdeburg, Germany\\
Tel: (0049) 391 671 8841\\
Fax: (0049) 391 671 1217\\
E-mail: Johannes.Richter@Physik.Uni-Magdeburg.DE

\clearpage

\renewcommand\baselinestretch {1.2}
\large\normalsize

Much work has been done
since the famous paper by Lieb, Schultz and Mattis \cite{lsm}
to derive exact results for thermodynamics and spin correlations
of one-dimensional spin-$\frac{1}{2}$ $XY$ models.
Much less exact results were obtained for random versions
of spin-$\frac{1}{2}$ $XY$ chains.
One can mention here a group of papers
dealing with random spin-$\frac{1}{2}$ $XY$ models using well-known 
Dyson's and Lloyd's 
models of disorder
\cite{s,n,dr}.
Recently the interest in random spin-$\frac{1}{2}$ $XY$ chains
has been noticebly increased
since they provide a laboratory for investigation
of generic features
of quantum phase transitions in disordered systems.
As an example we refer to papers on renormalization group \cite{f}
and numerical \cite{yr} studies
on random spin-$\frac{1}{2}$ transverse Ising chain.
 
In the present paper
we continue the study started in Ref. \cite{dr}
that concerns the spin-$\frac{1}{2}$ isotropic $XY$ chain
with random Lorentzian exchange coupling $J_n$
and a transverse field $\Omega_n$ that depends linearly
on the surrounding exchange couplings 
$J_{n-1}$ and $J_n$.
Obviously, due to the relation between the transverse field and
the random exchange couplings that is a model of correlated disorder.
The Jordan-Wigner method \cite{lsm} and the method elaborated by
John and Schreiber \cite{js}
permitted to derive exactly the averaged density of
states for such a model and as a result to study its thermodynamic
properties. Apparently the most interesting result of introducing
the correlated disorder is the appearance of the nonzero averaged
transverse magnetization at zero averaged transverse field.
Later this effect was checked numerically \cite{gc,dk}.
In the present communication we shall extend the model
introducing additional Dzyaloshinskii-Moriya interspin interaction.
Spin-$\frac{1}{2}$ $XY$ chains with Dzyaloshinskii-Moriya interaction
were studied in several papers \cite{kts,scg,dm,dkv,dv}
in which it was shown that they exhibit interesting
thermodynamic and dynamic properties, which may be of interest for the
understanding 
of the properties of 
some quasi-one-dimensional compounds
(e.g. CsCuCl$_3$).
It will be shown below that the Dzyaloshinskii-Moriya
interaction may influence 
in specific manner 
the thermodynamic properties of a magnetic chain
conditioned by correlated disorder.

Hereafter we consider isotropic $XY$ chain
in a magnetic field along $z$ axis
consisting of $N$ spins $\frac{1}{2}$. 
The Hamiltonian is defined by
\begin{eqnarray}
H=\sum_{n=1}^N\Omega_ns_n^z
+\sum_{n=1}^NJ_n(s_n^xs_{n+1}^x+s_n^ys_{n+1}^y)
+\sum_{n=1}^ND_n(s_n^xs_{n+1}^y-s_n^ys_{n+1}^x),
\nonumber\\
s_{n+N}^{\alpha}=s_{n}^{\alpha}.
\end{eqnarray}
Besides the exchange coupling $J_n$ between the neighbouring sites
$n$ and $n+1$ an additional Dzyaloshinskii-Moriya interaction $D_n$
between these sites is introduced,
i.e. a more general case
than in Ref. \cite{dr} is considered.

In what follows we consider two models.\\
{\it Model (i) --- } 
We assume 
the Dzyaloshinskii-Moriya interaction to be ordered,
i.e. $D_n=D$,
whereas the exchange couplings
$J_n$ are independent random Lorentzian variables 
with the probability distribution
\begin{eqnarray}
p(J_n)
=\frac{1}{\pi}
\frac{\Gamma}{(J_n-J_0)^2+\Gamma^2}.
\end{eqnarray}
The on-site transverse fields are determined by the formula
\begin{eqnarray}
\Omega_n-\Omega_0
=\frac{a}{2}(J_{n-1}+J_n-2J_0)
\end{eqnarray}
where $a$ is real and $\mid a\mid\ge 1$.
Note that after putting $D=0$ one obtains the model
considered in Ref. \cite{dr}.\\
{\it Model (ii) --- } 
We assume the exchange coupling to be ordered,
i.e. $J_n=J$, 
whereas
the $D_n$ are 
independent random Lorentzian variables with the probability
distribution
\begin{eqnarray}
p(D_n)
=\frac{1}{\pi}
\frac{\Gamma}{(D_n-D_0)^2+\Gamma^2}.
\end{eqnarray}
The on-site transverse fields are determined by the formula
\begin{eqnarray}
\Omega_n-\Omega_0
=\frac{a}{2}(D_{n-1}+D_n-2D_0)
\end{eqnarray}
where $a$ is real and $\mid a\mid\ge 1$.

With the help of the Jordan-Wigner transformation
the Hamiltonian (1) can be rewritten as a Hamiltonian of
non-interacting spinless fermions
\begin{eqnarray}
H=\sum_{n=1}^N\Omega_n
\left(c_n^+c_n-\frac{1}{2}\right)
+\sum_{n=1}^N\left(\frac{J_n+{\mbox{i}}D_n}{2}c_n^+c_{n+1}-
\frac{J_n-{\mbox{i}}D_n}{2}c_nc_{n+1}^+\right)
\end{eqnarray}
with cyclic boundary conditions.
We omitted in (6) the boundary term
that is not essential for the calculation of the thermodynamic properties
\cite{sm}.
Let us introduce the retarded and advanced temperature
double-time Green
functions
$G_{nm}^{\mp}(t)=\mp{\mbox{i}}\theta(\pm t)
\langle\{c_n(t),c_m^+\}\rangle$,
$G_{nm}^{\mp}(t)=\frac{1}{2\pi}\int_{-\infty}^{\infty}
{\mbox{d}}\omega{\mbox{e}}^{-{\mbox{i}}\omega t}
G_{nm}^{\mp}(\omega\pm{\mbox{i}}\epsilon)$
that satisfy the set of equations
\begin{eqnarray}
(\omega\pm{\mbox{i}}\epsilon-\Omega_n)
G_{nm}^{\mp}(\omega\pm{\mbox{i}}\epsilon)
\nonumber\\
-\left[
\frac{J_{n-1}-{\mbox{i}}D_{n-1}}{2}
G_{n-1,m}^{\mp}(\omega\pm{\mbox{i}}\epsilon)
+\frac{J_n+{\mbox{i}}D_n}{2}
G_{n+1,m}^{\mp}(\omega\pm{\mbox{i}}\epsilon)
\right]
=\delta_{nm}.
\end{eqnarray}
Our task is to evaluate the random-averaged Green functions since
they yield the random-averaged density of states through the relation
\begin{eqnarray}
\overline{\rho(E)}
=\mp\frac{1}{\pi}
{\mbox{Im}}
\overline{G_{nn}^{\mp}(E\pm{\mbox{i}}\epsilon)}.
\end{eqnarray}
Having the independent Lorentzian random variables
one may try to perform the random averaging of Eq. (7)
with the help of contour integrals. However one must know
the positions of the singularities
of the Green functions
in the planes of complex random variables.
The latter information can be derived for the defined models on
the basis of the Gershgorin criterion \cite{g}.

Consider at first spin model (i) described by Eqs. (1) - (3).
Suppose that exchange couplings $J_n$
(and hence the transverse fields $\Omega_n$)
are complex variables.
As it follows from (7) the singularities of
the matrix
${\bf{G}}^{\mp}
=\mid\mid G_{nm}^{\mp}(\omega\pm{\mbox{i}}\epsilon)\mid\mid$
are determined by the zeros of the determinant of the matrix
${\bf{A}}\pm{\mbox{i}}{\bf{B}}^{\mp}$
where $\bf A$ and ${\bf{B}}^{\mp}$ are the Hermitian matrices given by
\begin{eqnarray}
{\bf{A}}=
\left(
\begin{array}{ccccc}
\omega-{\mbox{Re}}\Omega_1 &
-\frac{1}{2}{\mbox{Re}}J_1-\frac{{\mbox{i}}}{2}D &
0 &
\ldots &
-\frac{1}{2}{\mbox{Re}}J_N+\frac{{\mbox{i}}}{2}D \\
-\frac{1}{2}{\mbox{Re}}J_1+\frac{{\mbox{i}}}{2}D &
\omega-{\mbox{Re}}\Omega_2 &
-\frac{1}{2}{\mbox{Re}}J_2-\frac{{\mbox{i}}}{2}D &
\ldots &
0 \\
\vdots &
\vdots &
\vdots &
\vdots &
\vdots \\
-\frac{1}{2}{\mbox{Re}}J_N-\frac{{\mbox{i}}}{2}D &
0 &
0 &
\cdots &
\omega-{\mbox{Re}}\Omega_N
\end{array}
\right)
\end{eqnarray}
and
\begin{eqnarray}
{\bf{B}}^{\mp}=
\left(
\begin{array}{ccccc}
\epsilon\mp{\mbox{Im}}\Omega_1 &
\mp\frac{1}{2}{\mbox{Im}}J_1 &
0 &
\ldots &
\mp\frac{1}{2}{\mbox{Im}}J_N \\
\mp\frac{1}{2}{\mbox{Im}}J_1 &
\epsilon\mp{\mbox{Im}}\Omega_2 &
\mp\frac{1}{2}{\mbox{Im}}J_2 &
\ldots &
0 \\
\vdots &
\vdots &
\vdots &
\vdots &
\vdots \\
\mp\frac{1}{2}{\mbox{Im}}J_N &
0 &
0 &
\cdots &
\epsilon\mp{\mbox{Im}}\Omega_N
\end{array}
\right),
\end{eqnarray}
respectively.
John and Schreiber noticed that if all eigenvalues of
${\bf{B}}^{\mp}$ are positive then
${\mbox{det}}({\bf{A}}\pm{\mbox{i}}{\bf{B}}^{\mp})\ne 0$
\cite{js}.
On the other hand for any eigenvalue $\lambda$ of the matrix
${\bf{B}}^{\mp}$ (10)
the Gershgorin criterion
after making use of Eq. (3)
guarantees that at least
one of the inequalities
\begin{eqnarray}
\left\vert\epsilon\mp
\frac{a}{2}
\left(
{\mbox{Im}}J_{n-1}
+{\mbox{Im}}J_n
\right)
-\lambda\right\vert
\le\frac{1}{2}\left\vert{\mbox{Im}}J_{n-1}\right\vert
+\frac{1}{2}\left\vert{\mbox{Im}}J_n\right\vert,
\;\;\;\mid a\mid\ge 1,
\;\;\;n=1,\ldots,N
\end{eqnarray}
is satisfied.
From (11) it immediately follows that the retarded (advanced)
Green function does not have poles for
${\mbox{Im}}J_n<0$
(${\mbox{Im}}J_n>0$)
if $a\ge1$
and for
${\mbox{Im}}J_n>0$
(${\mbox{Im}}J_n<0$)
if $a\le-1$.
Noting that
$\overline{F(\ldots,\Omega_n,J_n,\ldots)}
=F(\ldots,\Omega_0-{\mbox{i}}a\Gamma,J_0-{\mbox{i}}\Gamma,\ldots)$
if $F(\ldots,\Omega_n,J_n,\ldots)$ does not have poles
in lower half-planes
$J_n$ and
$\overline{F(\ldots,\Omega_n,J_n,\ldots)}
=F(\ldots,\Omega_0+{\mbox{i}}a\Gamma,J_0+{\mbox{i}}\Gamma,\ldots)$
if $F(\ldots,\Omega_n,J_n,\ldots)$ does not have poles
in upper half-planes
$J_n$ one finds the following result of
averaging the set
of equations (7)
\begin{eqnarray}
(\omega-\Omega_0\pm{\mbox{i}}\mid a\mid\Gamma)
\overline{G_{nm}^{\mp}(\omega)}
\nonumber\\
-\left[
\frac{J_0-{\mbox{i}}D\mp{\mbox{i}}\;{\mbox{sgn}}(a)\;\Gamma}{2}
\overline{G_{n-1,m}^{\mp}(\omega)}
+\frac{J_0+{\mbox{i}}D\mp{\mbox{i}}\;{\mbox{sgn}}(a)\;\Gamma}{2}
\overline{G_{n+1,m}^{\mp}(\omega)}
\right]
=\delta_{nm}.
\end{eqnarray}
The obtained equations (12) possess translational symmetry and
proceeding further in standard manner
one obtains
\begin{eqnarray}
\overline{\rho(E)}
=\frac{1}{\pi}
\sqrt{\frac{\sqrt{A^2+B^2}-A}{2(A^2+B^2)}},
\nonumber\\
A=(E-\Omega_0)^2+(1-\mid a\mid^2)\Gamma^2-J_0^2-D^2,
\nonumber\\
B=2\Gamma [\mid a\mid(E-\Omega_0)+{\mbox{sgn}}(a)\;J_0].
\end{eqnarray}

Consider now spin model (ii) described by Eqs. (1), (4), (5). 
Certainly we may repeat the whole calculation once more obtaining
as a result $\overline{\rho(E)}$ for this model.
However there is a relationship between the models
(i) and (ii) that immediately yields
thermodynamics of the latter model if it is known for the former one.
Namely, consider the following rotations of spin axes around $z$ axis
\begin{eqnarray}
\left(s_n^{\alpha}\right)^{\prime}
=\exp\left[-{\mbox{i}}\frac{\pi(n-1)}{2}s_n^z\right]
\;s_n^{\alpha}\;
\exp\left[{\mbox{i}}\frac{\pi(n-1)}{2}s_n^z\right].
\end{eqnarray}
One immediately finds that the Hamiltonian (1)
arises
as a result of transformations (14)
applied to a Hamiltonian of form (1),
however, with the exchange couplings $D_n$
and the Dzyaloshinskii-Moriya interactions $-J_n$.
Therefore it becomes evident
that the density of states (13)
after the replacement
$J_0\rightarrow D_0$,
$D^2\rightarrow J^2$
transforms into the density of states
for the model (ii).
Hence it is sufficient
in what follows to consider only the spin model (i) defined by (1) - (3).

Let us discuss the obtained density of magnon states (13).
It can be straightforwardly checked that (13)
covers in the particular case $D=0$
the result derived in Ref. \cite{dr}.
In the limit of diagonal disorder
$\Gamma\rightarrow 0$,
$\mid a\mid\Gamma=\gamma={\mbox{const}}$ 
$\;\;\;$
Eq. (13) reproduces the density of states for
spin-$\frac{1}{2}$ isotropic $XY$ chain
with Dzyaloshinskii-Moriya interaction
in a random Lorentzian transverse field with
the mean value $\Omega_0$ and
the width of distribution
$\gamma$ \cite{dv}.
The density of states (13) remains the same after the simultaneous change
of signs of $J_0$ and $a$; hereafter we choose $J_0>0$.

Let us remind how the density of states 
is influenced by correlated disorder in case of $D=0$ (for details see
\cite{dr}).
For $\mid a\mid\approx 1$ the disorder causes a smearing
out of mainly one edge of the magnon band (which one depends on the sign
of $a$). As a result we have
$\int_{-\infty}^{0}{\mbox{d}}E\overline{\rho(E)}
\ne
\int_{0}^{\infty}{\mbox{d}}E\overline{\rho(E)}$
at $\Omega_0=0$ that leads to 
the appearance of a nonzero averaged transverse
magnetization
$\overline{m_z}=-\frac{1}{2}
\int_{-\infty}^{\infty}{\mbox{d}}E\overline{\rho(E)}
\tanh \frac{\beta E}{2}$
at zero averaged transverse field $\Omega_0$.
With an increase of $\mid a\mid$ the symmetry of the non-random case
is recovered, i.e., both edges of the magnon band become smeared out
in a symmetric way,
the numbers of states
$\int_{-\infty}^{0}{\mbox{d}}E\overline{\rho(E)}$
and
$\int_{0}^{\infty}{\mbox{d}}E\overline{\rho(E)}$
at $\Omega_0=0$
become equal to each other,
and $\overline{m_z}=0$ at $\Omega_0=0$.

Figs. 1a, 1b demonstrate the changes in the 
behaviour of the averaged density of states
$\overline{\rho(E)}$ versus $E-\Omega_0$ for $\Gamma=1$,
$a=\pm 1.01$, $J_0=1$
for three different strengths of the
Dzyaloshinskii-Moriya interaction
$D=0$, $D=1$, $D=2$. It can be seen that an additional
Dzyaloshinskii-Moriya interspin interaction
1) increases the width of the smoothed magnon band;
2) leads to the recovering of the symmetry with respect to the change
$E-\Omega_0\rightarrow -(E-\Omega_0)$.
Thus the increase of the Dzyaloshinskii-Moriya interaction
leads to the decrease of the nonzero value of
$\overline{m_z}$ at $\Omega_0=0$ (Figs. 1c, 1d).

In Fig. 2 we depicted the influence of an increase of the
averaged exchange coupling $J_0$ at fixed $D=0$.
Similarly to the previous case one observes an increasing of the band width,
however, in contrast to the previous case the density of states remains
not symmetric with respect to the change
$E-\Omega_0\rightarrow -(E-\Omega_0)$
(Figs. 2a, 2b)
and as a result the model exhibits
a noticeable nonzero value of
$\overline{m_z}$ at $\Omega_0=0$
(Figs. 2c, 2d).
The difference in the behaviour of the density of states
with increasing $D$ or $J_0$
is not surprising since $J_0$ and $D$ enter in a different way
into (13).

To summarize,
we have studied the 
spin-$\frac{1}{2}$ transverse isotropic $XY$ chain
in the presence of correlated Lorentzian 
disorder. 
Going beyond the results given in Ref. \cite{dr} we include in the
model the  
Dzyaloshinskii-Moriya interaction.
The asumption of correlated disorder allows the exact calculation of the
averaged density of states $\overline{\rho(E)}$. 
The exact formula (13) for $\overline{\rho(E)}$ is the main
result of the paper. Based on this formula one can calculate in a simple
way exactly the 
thermodynamic properties like entropy, specific heat, 
transverse magnetization  
and static transverse linear susceptibility 
(see for details \cite{dr}). In that
sense the presented random quantum spin model may serve as a reference model
to study the interplay of disorder and quantum effects. In particular, it
may be used to test approximations and/or calculations for finite systems.
As an example we present results for the density of states and the
transverse magnetization.
In particular, we find that the  
Dzyaloshinskii-Moriya interaction  may lead to a decrease of the nonzero
averaged transverse magnetization at zero averaged transverse field 
that appears due to correlated disorder. 
It is known \cite{kts,scg,dm,dkv}
that in the non-random case 
the Dzyaloshinskii-Moriya interaction leads to spectacular changes in the spin
correlations and their dynamics.
However, the rigorous consideration of correlated disorder in this paper 
is restricted to thermodynamic quantities based on the density of states.
The effect of the Dzyaloshinskii-Moriya interaction on the 
spin correlations and their dynamics in the presence of correlated 
disorder may be studied numerically \cite{dk2}.
\\[15pt]
The authors thank the DFG (project Ri615/1-2)
for support of the present study.
The paper was presented at the XXth IUPAP International Conference 
on Statistical Physics (Paris, 1998).
O.D. is grateful to the Grant Committee for a financial support
for attending the Conference.
He also thanks the University of Magdeburg for hospitality
in the autumn of 1998 when the paper was finished.

\clearpage

\begin{figure}
\epsfxsize=100mm
\epsfbox{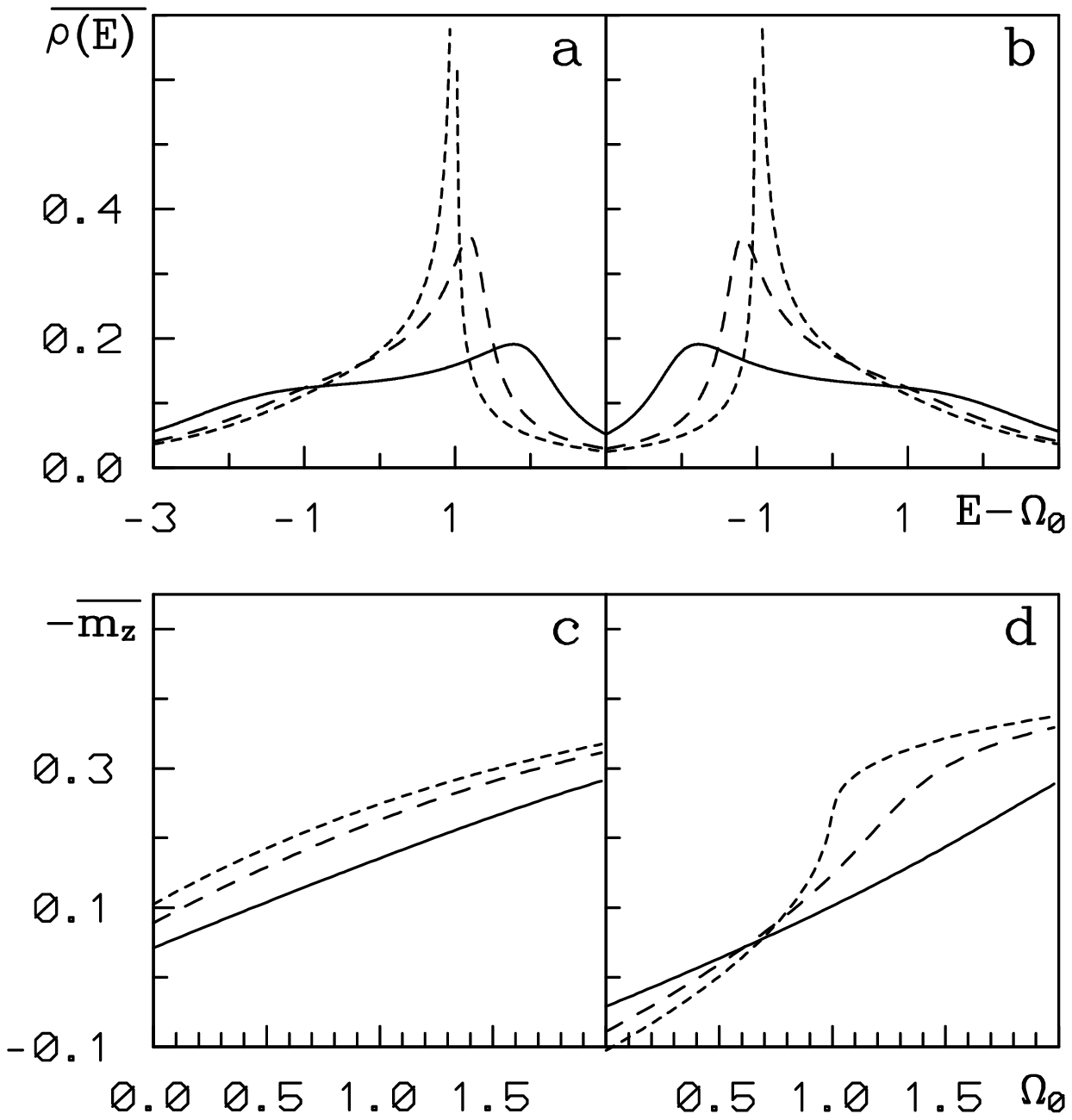}
\vspace{60mm}
\caption{
The density of states (described by Eq. (13)) (Figs. 1a, 1b)
and the transverse magnetization $-\overline{m_z}$ versus $\Omega_0$
at $\beta=1000$ (Figs. 1c, 1d) at fixed 
$J_0=1$, $\Gamma=1$ and $a=-1.01$ (Figs. 1a, 1c) 
or $a=1.01$ (Figs. 1b, 1d).
The short-dashed curves correspond to  $D=0$,
long-dashed curves to $D=1$ and the 
solid curves to $D=2$.
}
\end{figure}

\clearpage

\begin{figure}
\epsfxsize=100mm
\epsfbox{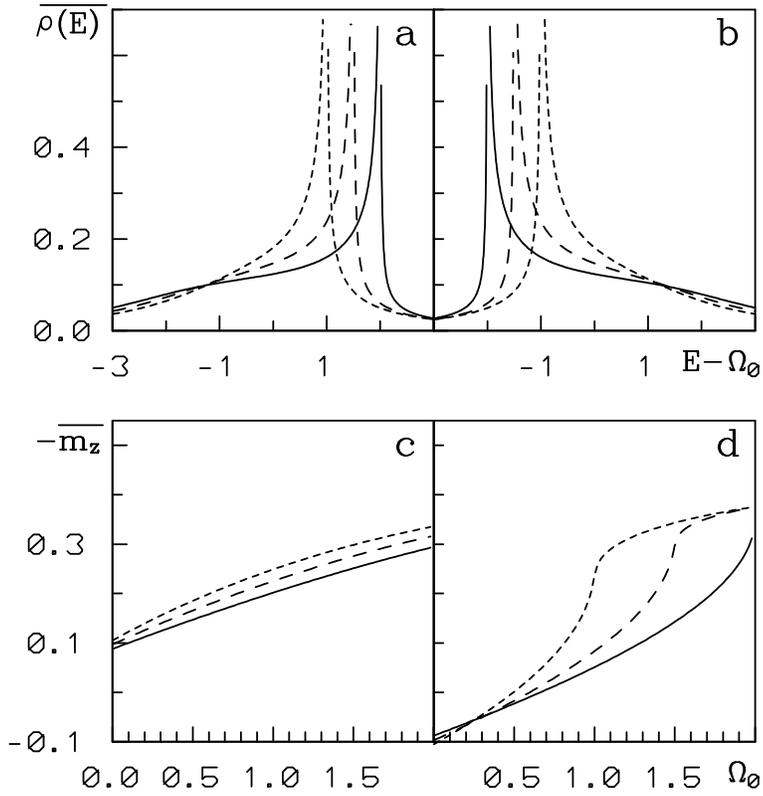}
\vspace{60mm}
\caption{
The density of states (described by Eq. (13)) (Figs. 2a, 2b)
and the transverse magnetization $-\overline{m_z}$ versus $\Omega_0$
at $\beta=1000$ (Figs. 2c, 2d) at fixed 
$D=0$, $\Gamma=1$ and $a=-1.01$ (Figs. 2a, 2c) 
or $a=1.01$ (Figs. 2b, 2d).
The short-dashed curves correspond to  $J_0=1$,
long-dashed curves to $J_0=1.5$ and the 
solid curves to $J_0=2$.}
\end{figure}

\clearpage
\noindent
{\bf List of figure captions}

\vspace{1.25cm}

FIG. 1.
The density of states (described by Eq. (13)) (Figs. 1a, 1b)
and the transverse magnetization $-\overline{m_z}$ versus $\Omega_0$
at $\beta=1000$ (Figs. 1c, 1d) at fixed 
$J_0=1$, $\Gamma=1$ and $a=-1.01$ (Figs. 1a, 1c) 
or $a=1.01$ (Figs. 1b, 1d).
The short-dashed curves correspond to  $D=0$,
long-dashed curves to $D=1$ and the 
solid curves to $D=2$.

\vspace{1.25cm}

FIG. 2.
The density of states (described by Eq. (13)) (Figs. 2a, 2b)
and the transverse magnetization $-\overline{m_z}$ versus $\Omega_0$
at $\beta=1000$ (Figs. 2c, 2d) at fixed 
$D=0$, $\Gamma=1$ and $a=-1.01$ (Figs. 2a, 2c) 
or $a=1.01$ (Figs. 2b, 2d).
The short-dashed curves correspond to  $J_0=1$,
long-dashed curves to $J_0=1.5$ and the 
solid curves to $J_0=2$.

\end{document}